\documentclass[prb,twocolumn,showpacs,preprintnumbers,amsmath,amssymb]{revtex4}

\usepackage{epsfig}
\usepackage{graphicx}
\usepackage{dcolumn}
\usepackage{bm}

\newcommand{\la}{\langle}
\newcommand{\ra}{\rangle}

\newcommand{\e}{\epsilon}

\newcommand{\vf}{v_{\rm F}}

\newcommand{\ef}{\epsilon_{\rm F}}
\newcommand{\ek}{\epsilon_{\bm k}}

\newcommand{\kf}{k_{\rm F}}

\newcommand{\Hkin}{{\cal H}_{\rm kin}}
\newcommand{\Hex}{{\cal H}_{\rm I}}
\newcommand{\pk}{\phi_{\bm k}}

\newcommand{\pa}{\psi_a}
\newcommand{\pb}{\psi_b}

\newcommand{\eo}{\epsilon_0}

\newcommand{\kc}{k_c}
\newcommand{\ku}{k_{\uparrow}}
\newcommand{\kd}{k_{\downarrow}}

\newcommand{\Se}{\Sigma ( \omega )}
\newcommand{\So}{\Sigma_0}

\begin{document}

\title{Coulomb Interactions and Ferromagnetism in Pure and Doped Graphene}
\author{N. M. R. Peres$^{1,2}$, F. Guinea$^{1,3}$, and A. H. Castro Neto$^1$}

\affiliation{$^1$Department of Physics, Boston University, 590 
Commonwealth Avenue, Boston, MA 02215,USA}
\affiliation{$^2$Center of Physics and Department  of Physics,
Universidade do Minho, P-4710-057, Braga, Portugal}
\affiliation{$^3$Instituto de Ciencia de Materiales de Madrid, CSIC,
 Cantoblanco E28049 Madrid, Spain}

\begin{abstract}
We study the presence of ferromagnetism in the phase diagram of  the
two-dimensional honeycomb lattice close to 
half-filling (graphene) as a function of the strength of the Coulomb interaction and doping. 
We show that exchange interactions between Dirac fermions can stabilize a
ferromagnetic phase at low doping when the coupling is sufficiently large. 
In clean systems the zero temperature phase diagram shows both first order
and second order transition lines and two distinct ferromagnetic phases: 
one phase with only one type of carriers (either electrons or holes) 
and another with two types of carriers (electrons and holes). 
Using the coherent potential approximation we argue that disorder further 
stabilizes the ferromagnetic phase. This work should estimulate Monte Carlo
calculations in graphene dealing with the long-range nature of the 
Coulomb potencial.  
\end{abstract}

\maketitle
\section{Introduction}

The ferromagnetic instability due to the exchange interaction in a three
dimensional (3D) electron gas attracted attention since the early days of
quantum mechanics \cite{Bloch} 
and has been studied in great detail \cite{Iwamoto62,Misawa65}. 
Recent Monte Carlo calculations \cite{Ceperley78,Ceperley80} have confirmed the presence of 
ferromagnetism in the phase diagram of the 3D electron gas at
low doping. Similar studies have also suggested the existence of a
ferromagnetic phase 
in the diluted two dimensional (2D) electron gas \cite{Attaccalite02} with a 
first order transition from a paramagnetic phase to a ferromagnetic phase with full polarization. 
As the electron density is reduced, electron-electron interactions become
stronger and dynamical screening disappear. At the extreme limit of zero
density the electron gas should crystalize into a Wigner solid where the
electrons feel the unscreened Coulomb interaction. The elusive ferromagnetic
phase of the electron gas lurks between the Wigner crystal and the Fermi 
liquid state that exists at higher doping  when electron-electron interactions are fully screened \cite{Attaccalite02,ceperley_nature}. 
  
  \begin{figure}[htf]
\begin{center}
\includegraphics*[width=8cm]{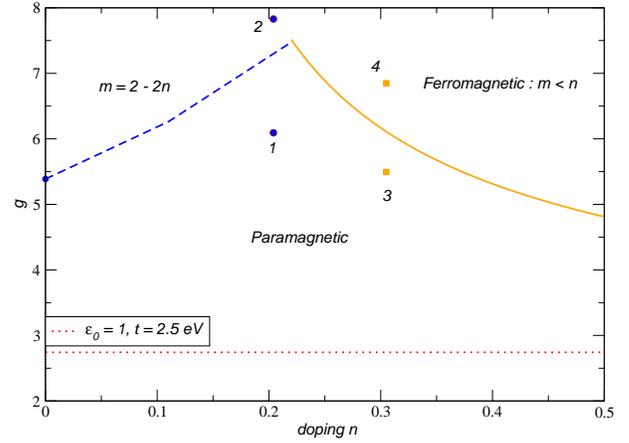}
\end{center}
\caption{
\label{phase_diagram}
Zero temperature phase diagram of a clean
graphene plane as a function of the coupling constant $g$, eq.~(\ref{g}),
and doping away from half-filling. The dashed line corresponds to a
first order, and the continuous line a second order
phase transition between the paramagnetic and ferromagnetic phases. 
The dotted curve corresponds to the value of $g$ with $\eo=1$ and 
a Dirac-Fermi velocity of $\hbar \vf = 5.7$ eV $\AA$, as defined by
Eq.(\ref{g}). The points labeled 1-4
in the figure are discussed ahead in the text in connection with Fig. \ref{phase_diag}.}
\end{figure}
  
In recent years, the experimental search for the ferromagnetic phase of the
diluted electron gas has not been succesful \cite{ceperley_nature,fisk,fiasco}. 
Nevertheless, there has been strong experimental indications on the existence
of ferromagnetism in highly disordered graphite samples \cite{esquinazi,disorder}. 
The origin of this phase is still unclear, and a number of different
mechanisms have been proposed \cite{OS91,H01,Letal04,Vetal05}. Nevertheless,
there is no final word on the origin of ferromagnetism in graphite. 
Graphite is a layered material made out of graphene layers (a honeycomb
lattice with one electron per $\pi$ orbital, that is, a half-filled band). 
The traditional view of graphite based on band-structure calculations {\it
  assumes} coherent hopping between graphene layers, and describes graphite 
as a low density metal with almost compensated electron and hole pockets, 
with $10^{-4}$ to $10^{-5}$ electrons per Carbon \cite{BCP88}. This
traditional picture, however, completely disregards the strong and unscreened 
interactions between electrons that should exist at low densities. 
In fact, recent experiments in true 2D graphene systems
\cite{Netal04,Zetal04,Zetal05,Betal04} 
show that 
electron-electron interactions and disorder have to be taken into account in
order to obtain a fully consistent picture of graphene \cite{nuno}. 
Recent theoretical results \cite{nuno} raise questions on the wisdom of
thinking of strongly correlated layered system such as graphite, as truly
3D. The claim is that the full 2D nature of graphene
has to be taken into account before graphene planes are coupled by weak 
van der Waals interactions in order to form the 3D solid. 

One of the most striking features of the electronic structure of perfect graphene 
planes is the linear relationship between the electronic energy, $E_{{\bm k}}$, with the 
two-dimensional momentum, ${\bm k} =(k_x,k_y)$, that is: $\epsilon({\bm k}) = 
\pm\hbar \vf |{\bm k}|$,
where $\vf$ is the Dirac-Fermi velocity. This singular dispersion relation is a
direct consequence of the honeycomb lattice structure that can be seen
as two interpenetrating triangular sublattices. In ordinary metals and semiconductors
the electronic energy and momentum are related quadratically via the so-called 
effective mass, $m^*$, ($E_{{\bm k}} = \hbar^2 {\bm k}^2/(2 m^*)$), that controls 
much of their physical properties. Because of the linear dispersion relation,
the effective mass in graphene is zero, leading to an unusual electrodynamics.
In fact, graphene can be described mathematically 
by the 2D Dirac equation, whose elementary excitations are 
particles and holes (or anti-particles), in close analogy with systems in particle 
physics. In a perfect graphene sheet the chemical potential crosses the Dirac point
and, because of the dimensionality, the electronic density of states 
vanishes at the Fermi energy. The vanishing of the effective mass or density of
states has profound consequences.  It has been shown, for instance, that the Coulomb 
interaction, unlike in an ordinary metal, remains unscreened and gives rise to an 
inverse quasi-particle lifetime  that increases linearly with energy or temperature 
\cite{GGV96}, in contrast with the usual metallic Fermi liquid paradigm, where the 
inverse lifetime increases quadratically with energy. 

As mentioned above, its is well known that direct exchange interactions can
lead to a ferromagnetic instability in a dilute electron gas \cite{Bloch,S47}.
In this work we generalize the analysis of the exchange
instability of the electron gas to pure and doped 2D graphene
sheets. Although pure graphene should be a half-filled system,
we have recently shown \cite{nuno} that extended defects 
such as dislocations, disclinations, edges, and micro-cracks can lead
to the phenomenon of self-doping where charge is transfered
to/from defects to the bulk in the presence of particle-hole asymmetry. 
The extended defects are unavoidable in graphene  because there
can be no long-range positional Carbon order at finite temperatures in 2D
(the Hohenberg-Mermin-Wagner theorem). Furthermore, we have also
shown that although extended defects lead to self-doping, they do not
change the transport and electronic properties. Life-time effects are
actually introduced by localized disorder such as vacancies and ad-atoms.
Thus, we have also considered the influence of disorder in the 
generation of ferromagnetism.  It is worth noting that the possibility of other instabilities
in a graphene plane, related to the Coulomb interaction have also been studied in the 
literature \cite{K01,Getal02}. The nature of the exchange instability in a
system with many bands is also interesting on its own right \cite{Getal02b},
and it has not been studied extensively. Furthermore, graphene is the basic
material for the synthesis of other compounds with sp$^2$ bonding: graphite
is obtained by the stacking of graphene planes, Carbon nanotubes are synthesized
by the wrapping of graphene along certain directions, and fullerenes "buckyballs" are
generated from graphene by the creation of topological defects with five and seven
fold symmetry. Therefore, the understanding of the ferromagnetic instability in
graphene can have impact on a large class of systems. Finally, we also 
mention that a simple analysis using the standard Stoner criterium for ferromagnetism 
fails in graphene, as the density of states of undoped graphene vanishes at
the Fermi level \cite{PAB04}. 

The electron-electron interaction in graphene can lead to other instabilities
at low temperatures, in addition to the ferromagnetic phase considered
here. A local on site repulsive term can lead to an antiferromagnetic phase,
when its value exceeds a critical threshold\cite{ST92,PAB04}. In
the following, we will concentrate on the role of the ferromagnetic exchange
instability, which, as already mentioned, is important in electronic systems
with a low density of carriers, and which has not been considered in the
literature so far.

Our main results can be summarized by the zero-temperature phase 
diagram $g${\it versus} $n$ 
(where $n$ is the doping away from half-filling) shown in
Fig. \ref{phase_diagram}. The strength of the electron-electron interactions
in graphene is parameterized by the dimensionless coupling constant, $g$, defined as:
\begin{eqnarray}
g = \frac{e^2/\epsilon_0}{\hbar v_{\rm F}} \, ,
\label{g}
\end{eqnarray}
where  $e$ is the charge of the electron, and $\epsilon_0$ the dielectric constant
of the system. Notice that $g$ is exactly the ratio between the Coulomb to
the kinetic energy of the electron system. This coupling constant replaces the
well-known parameter $r_s \sim (e^2/\epsilon_0)/[\hbar^2 k_F/m^*]$ 
of the non-relativistic electron gas (where $k_F$ is the Fermi momentum). 
In the pure compound ($n = 0$) the paramagnetic-ferromagnetic transition
is of first order with partial polarization and occurs at a critical 
value of $g = g_c \approx 5.3$. As the
doping is increased, the ferromagnetic transition is suppressed (a larger
value of $g_c$ is required) up to around $n \approx 0.2$ where the
first order line ends at a tri-critical point a line of second order transitions
emerges with a fully polarized ferromagnetic phase. A unique feature of
the ferromagnetism in these systems, unlike the ordinary 2D and 3D electron
gases, is the fact that there are two types of ferromagnetic phases, one that
has only one type of carrier (either electron or hole) and a second phase
with two types of carriers (electrons and holes).

The paper is organized as follows: in the next section we present the
model for a graphene plane in the continuum limit taken into account
the Dirac fermion spectrum and the long-range Coulomb interactions;
in Section \ref{section_exchange} we discuss the exchange energy for graphene
through a variational wavefunction calculation in three different situations:
Dirac fermions without a gap; Dirac fermions with a gap; and Dirac fermions
with disorder treated within the coherent potential approximation (CPA) 
approximation; Section \ref{conclusions}
contains our conclusions. We also have included two appendixes with the
details of the calculations.

\section{The model for a graphene layer}
\label{2dDirac}

The valence and conducting bands in graphene are formed by 
Carbon $\pi$ orbitals which are arranged in an honeycomb
lattice (a non-Bravais lattice). 
The extrema of these bands lie at the $\Gamma$ point and at the
two inequivalent corners of the hexagonal Brillouin Zone. 
When the filling is close to one
electron per Carbon atom, the Fermi energy lies close to the corners. 
Near these
points, a standard long wavelength expansion gives for the kinetic
part of the Hamiltonian the expression,
\begin{equation}
\Hkin ( \bm k ) \equiv \hbar \vf \left( 
\begin{array}{cc} 0 &k_x + i k_y \\ k_y - i k_y &0
  \end{array} \right)\,, \label{hkin}
\end{equation}
which leads to the dispersion relation,
\begin{equation}
\e({\bm k}) = \pm \hbar \vf \vert {\bm k} \vert\, .
\label{dispersion}
\end{equation} 
In a tight-binding description of the graphene plane with
nearest neighbor hopping energy $t$ the Dirac-Fermi
velocity is given by:
\begin{eqnarray}
\hbar v_{\rm F} = \frac{3}{2} t a
\end{eqnarray}
where $a$ is the Carbon-Carbon distance ($t \approx 2.5$ eV and $a = 1.42 \, \AA$) \cite{BCP88}.
The eigenstates of (\ref{hkin}) can be written as:
\begin{eqnarray}
\Psi_{\bm k,\alpha,\sigma} ( \bm r ) &\equiv& \left( \begin{array}{c} 
\pa ( \bm r ) \\ \pb ( \bm r ) 
\end{array} \right) \chi_\sigma \, ,
\nonumber
\\
&=& \frac{e^{i \bm k \cdot\bm r}}{\sqrt{2}} \left( \begin{array}{c} e^{i \pk / 2} 
    \\ \alpha e^{- i \pk / 2} 
\end{array} \right) \chi_\sigma\,,
\label{wavefunction}
    \end{eqnarray} 
where $a$ and $b$ label the two sublattices of the honeycomb lattice, 
$\pk = \arctan ( k_y / k_x )$ is a phase factor, $\alpha = \pm 1$ labels the
electron and hole-like bands, and 
$\chi_\sigma$ is the spin part of the wavefunction.
The dispersion and the wavefunctions are the solutions of the 2D
Dirac equation. This approach in the continuum requires the introduction
of a cut-off in momentum space, $k_c$, in such a way that all momenta, ${\bm k}$,
are defined such that:
$0 \le \vert \bm k \vert \le \kc\,$, where
$\kc$ is chosen so as to keep the number of states in the Brillouin zone is
fixed, that is,  $\pi \kc^2 = ( 2 \pi )^2 / A_0$, and $A_0$ is 
the area of the unit cell in the honeycomb lattice.

It is easy to show that with the dispersion given in (\ref{dispersion})
the single particle density of states, $\rho(E)$, vanishes linearly
with energy at the Dirac point, $\rho(E) \propto |E|$. In this case,
there is no electronic screening \cite{mele} and the electrons
interact through long-range Coulomb forces. 
The electron-electron interactions can be written 
in terms of the field operators, $\hat\Psi(\bm r)$, as:
\begin{equation}
\Hex = 
\frac{1}{2} \int d\bm r_1d\bm r_2 \hat\Psi^\dag(\bm r_1)
\hat\Psi^\dag(\bm r_2)V(\bm r_1-\bm r_2)
\hat\Psi(\bm r_2)
\hat\Psi(\bm r_1)\,,
\end{equation} 
where $V(\bm r) = e^2/(\epsilon_0 r)$ is the bare
Coulomb interaction. One can now expand the field
operators in the basis of states given in (\ref{wavefunction}),
that is, 
\begin{eqnarray}
\hat\Psi(\bm r) = \frac{1}{\sqrt{A}} \sum_{{\bm k},\alpha,\sigma} 
\Psi_{\bm k,\alpha,\sigma} ( \bm r ) a_{{\bm k},\alpha,\sigma}
\end{eqnarray} 
where $a_{{\bm k},\alpha,\sigma}$ ($a^{\dag}_{{\bm k},\alpha,\sigma}$) is the annihilation
(creation) operator for an electron with momentum ${\bm k}$, band $\alpha$, and spin $\sigma$ ($\sigma=
\uparrow,\downarrow$ and $A$ is the area of the system). In this case, the Coulomb interaction reads:
\begin{widetext}
\begin{eqnarray}
\Hex &=& \frac{2\pi e^2}{8 \eo A} \sum_{\bm k , \bm p, \bm q }\sum_{ 
\alpha_1,\ldots,\alpha_4}\sum_{\sigma,\sigma'} \frac 1{q}
[\alpha_2\alpha_3 e^{i[\phi^\ast(\bm p)-\phi(\bm p+\bm q)]}+1]
[\alpha_1\alpha_4 e^{i[\phi^\ast(\bm k)-\phi(\bm k+\bm q)]}+1]
a^\dag_{\bm k,\alpha_1,\sigma_1}a^\dag_{\bm p,\alpha_2,\sigma_2}
a_{\bm p+\bm q,\alpha_3,\sigma_2}a_{\bm k-\bm q,\alpha_4,\sigma_1}\, .
\nonumber\\
\label{exchange} 
\end{eqnarray}
\end{widetext}
It is easy to see that the Coulomb interaction induces scattering between
bands (inter-band) and also within each band (intra-band). Furthermore,
the $1/q$ dependence of the interaction (that comes from the Fourier
transform of the $1/r$ potential in 2D) provides an electron-electron
scattering that is stronger than in 3D, allowing for the possibility of
a ferromagnetic transition at weaker coupling. As in the case of  the Hund's
coupling in atomic systems, the spin polarized state is always preferred 
when long-range interactions are present since, by the Pauli's exclusion
principle, both kinetic and Coulomb energies are minimized simultaneously.
This should be contrast with ultra-short range interactions of the Hubbard
type that almost always benefit anti-ferromagnetic coupling via a kinetic
exchange mechanism. 

\section{Exchange energy of a graphene plane.}
\label{section_exchange}

In what follows we examine the required conditions for a ferromagnetic ground state in graphene. 
Our purpose in this work is not to obtain exact values for the critical couplings, that may required
more sophisticated approaches, but instead our aim is to show that a
ferromagnetic ground state in graphene is possible {\it in principle}. 
In order to study the ferromagnetic instability we use a
variational procedure that respects all the symmetries of the problem. We assume that: ({\it i}) the
ferromagnetic instability only affects states close to the Dirac points in the region at the edge
of the Brillouin zone (that is, long wavelength approximation is still valid); ({\it ii}) in the ferromagnetic
state the electronic bands are shifted rigidly (hence, self-energy effects such as Dirac-Fermi
velocity renormalizations are neglected); ({\it iii})  even when the bands
are shifted, and a finite density of states is produced at the Fermi energy,
the Coulomb interaction remains unscreened (this assumption is equivalent to
assume that the chemical potential shift is always small and that the
screening length is larger than the inter-particle distance); 
({\it iv}) the ferromagnetic state is uniform and translational
invariant. Besides considering the case of a gapless system, we have also studied the case
where a gap $\Delta$ opens in the Dirac spectrum (that is, when the dispersion relation becomes
$E_{{\bm k}} = \pm \sqrt{\Delta + \hbar^2 v^2_{\rm F}}$). The gapped case is interesting because
it allows the study of the crossover between the Dirac case when $\Delta =0$ to the standard
2D case with a finite effective mass $m^* \propto \Delta$
(see details ahead). We also briefly the discuss the
effects of disorder on the stabilization of the ferromagnetic state via a CPA approximation
in order to point out that disorder may be fundamental for the realization of a ferromagnetic
phase in graphite.

\subsection{Gapless system}
\subsubsection{Exchange energy. Inter- and intraband contributions.}
\begin{figure}
\begin{center}
\includegraphics*[width=9cm]{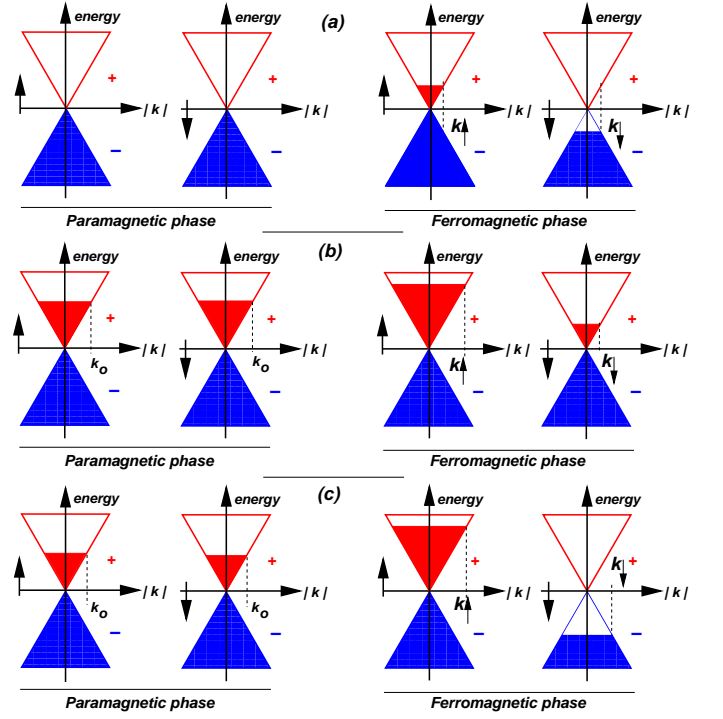}
\end{center}
\caption{Occupied and empty states
in the paramagnetic and ferromagnetic ground states of Dirac fermions {\bf (a)}
half-filling case; {\bf (b)} finite doping and one type of carrier in the ferromagnetic phase;
{\bf (c)}  finite doping and two types of carriers  
in the ferromagnetic phase.}
\label{cones}
\end{figure}

The possible ferromagnetic instability arises from the gain in exchange energy
when the system is polarized. A finite spin polarization, on the other hand,
leads to an increase in kinetic energy. Thus, there are two competing energies
in the problem: the exchange energy that is minimized by polarization and the
kinetic energy that is increased by it. The variational states that we consider
in our approach are Slater determinants of the wave-functions given
by (\ref{wavefunction}) in the configurations shown in Fig.\ref{cones}.

As function of the Fermi wave vector,
$\kf$, the kinetic energy of the unpolarized state is:
\begin{equation}
\la \Hkin \ra = K=- \frac{A}{3 \pi} \vf \hbar( \kc^3 - \kf^2 )\,,
\end{equation}
and the  exchange energy, for any doping,
as determined from Eq.(\ref{exchange})
can be written as
\begin{eqnarray}
E_{ex}=&-&\frac {A}{(2\pi)^2}\frac {e^2}{4\epsilon_0}
\sum_\sigma\sum_{\alpha_a,\alpha_b}\int_0^{2\pi}d\theta
\int kpdkdp\nonumber\\&&\frac {1+\alpha_a\alpha_b\cos\theta}{\vert
\bm k-\bm p\vert}
n_{\rm F}^{\sigma,\alpha_a}(\bm k)
n_{\rm F}^{\sigma,\alpha_b}(\bm p)\,, \label{exchange_integral}
\end{eqnarray}
where $n_{\rm F}^{\sigma,\alpha_a(\alpha_b)}(\bm k)$ 
is the Fermi occupation function, $a(b)$ is the 
band indice, and $\alpha_a , \alpha_b = \pm 1$.
 
In the ferromagnetic state the degeneracy of the spin states is lifted
and the Fermi momentum of the up and down spin states becomes 
$\ku$ and $\kd$, respectively. Depending on the values of $\kf , \ku$ and
$\kd$, we can define the three cases shown in Fig. \ref{cones}. 
For a doping, $\delta$ per unit area, the number of electrons per Carbon 
away from half-filling, $n$, can
be written as: 
\begin{eqnarray}
n=\delta A_0 \, .
\end{eqnarray}  
Because of the different values of $\ku$ and
$\kd$ the system acquires a spin magnetization, $\mu = g_s \mu_B m$, where 
$g_s \approx 2$ is the electron gyromagnetic factor, $\mu_B$ is the Bohr
magneton, and $m= s \, A_0$ with 
$s = n_{\uparrow}-n_{\downarrow}$, is the spin polarization. Notice that the
maximum polarization allowed is $m =  2 - 2n$ since each added (subtracted)
electron leads to a doubly (empty) Carbon $\pi$ orbital. 

 The total exchange energy, eq.(\ref{exchange_integral}), can be split into intra- and inter-band
  contributions. In many band systems where the different bands arise from
  different atomic orbitals, the overlap integral between Bloch states
  corresponding to different bands can be neglected,
  and, consequently, there are no inter-band contributions to the exchange
  energy. An analogous effect arises when the different bands are localized
  at different sites of the lattice, as in the gapful case to be considered
  below. There are also situations where the
  different bands arise from the same orbitals at the same sites, but their
  phases in a region much larger than the unit cell are such that the overlap integral
  vanishes. This is the case for the two different Dirac cones which can be
  defined in the honeycomb lattice. We do not need to include in
  eq.(\ref{exchange_integral}) terms due to interactions between electrons
  near different Dirac points of the Brillouin Zone.

The case studied here, where the overlap between Bloch states in different
bands cannot be neglected, and a corresponding term in the exchange energy
has to the included is generic to narrow gap semiconductors, and this term may be
important in lightly doped materials. 

It is worth noting that these inter-band exchange effect arise from the non local nature of the exchange
interaction. They cannot be studied when the exchange energy is approximated by a local term which only
depends on the total charge density.

\subsubsection{Undoped case: $n=0$}

The Fermi level in the paramagnetic case  is at $\ef = 0$, and the
bands are half-filled. Then, in the paramagnetic state one has $\ku = \kd$. 
When the system polarizes the magnetization is such that $\ku =
\sqrt{2 \pi s}$ and the change in energy relative to the paramagnetic
state is given by:
\begin{eqnarray}
\Delta E &=& \Delta K + \Delta E_{\rm ex} = \frac {A_0}{3 \pi} \hbar \vf \ku^3 
\nonumber\\
&-& \frac{A_0}{( 2 \pi )^2}\frac{e^2}{4 \eo}
\left [ 2 \ku^3  R_1(1)-4k_c \ku^2 R_0 \left( \frac{\ku}{\kc} \right) 
\right]\,,
\label{case_a}
\end{eqnarray}
where the functions $R_n(x)$ are defined in the Appendix \ref{contas}.
Unfortunately it is not possible to find an analytical expression 
(using elementary functions) for
the energy change as a function of the electron polarization $s = \ku^2/(2 \pi)$.
For $\ku \ll \kc$, the leading contribution comes from  the expansion
of function $R_0 ( x ) \approx -x \ln(x)$ for $x \ll 1$
(see Appendix \ref{contas}). Hence,
the exchange energy increases as the polarization increases, and a
ferromagnetic state with small magnetization is not favored. This effect can
be cast as a logarithmic renormalization of the Fermi energy, which reduces
the density of states near the Fermi level, and suppresses the tendency
toward ferromagnetism\cite{GGV99}.

At large magnetizations, $\kc^2/s \sim 1$, the kinetic energy contribution
tends to a term proportional to $\vf \kc^3$ and the exchange contribution
becomes negative and proportional to $- ( e^2 \kc^3 )/ \eo$. The exchange
term dominates, and the system undergoes a discontinuous transition to a
state with polarization of order unity when:
\begin{equation}
g_c = \frac{e^2}{\hbar \vf \eo} \ge \frac{16 \pi}{
6 R_1 ( 1 ) - 12 R_0 ( 1 )} \simeq 5.3\,,
\label{strong_coupling_a}
\end{equation}
which gives the critical coupling $g_c(n=0) \approx 5.3$ for the appearance
of ferromagnetism in the clean system, as shown in Fig.\ref{phase_diagram}.

\subsubsection{Doped case, $n \neq 0$, one type of carrier in the ferromagnetic phase}

In this case the doping, $\delta$, and magnetization, $s$, are such that 
$\kf =\sqrt{2\pi \delta}$ in the paramagnetic paramagnetic phase, 
and $\ku =\sqrt{2\pi(s+\delta)}$ and
$\kd =\sqrt{2\pi(s-\delta)}$ in the ferromagnetic phase. In this phase
there is only one type of carriers, either electrons or holes.
The change in energy between the paramagnetic and ferromagnetic
phase is:
\begin{eqnarray}
\Delta E &= &\Delta K + \Delta E_{\rm ex} = \frac {A_0}{6\pi}v_F\hbar
(\ku^3+\kd^3-2 \kf^3) \nonumber \\
&- &\frac{A_0}{(2 \pi )^2} \frac{e^2}{\eo}
 \left[\ku^3 R_1(1)+\kd^3 R_1(1)-2 \kf^3 R_1(1) \right.\nonumber\\
&+& 2 \kc \kd^2 R_2 \left( \frac{\kd}{\kc} \right)
 + 2 \kc \ku^2 R_2 \left( \frac{\ku}{\kc} \right)\nonumber\\
&-&4 \left.\kc \kf^2 R_2 \left( \frac{\kf}{\kc} \right) 
\right]\, .
\label{case_b}
\end{eqnarray}
The behavior of the energy change as a function of the spin polarization is
shown in left hand pannel in Fig.\ref{phase_diag} for points $1$ and $2$
of the phase diagram in Fig.\ref{phase_diagram}. Notice that the transition
between the paramagnetic phase (point $2$) to the ferromagnetic phase (point $1$)
is discontinuous with full polarization, $m=2 - 2n$. In this case analytical expansion when
$s \ll \delta$ is now possible. For $\kf , \ku , \kd \ll \kc$
the value of the exchange contribution is dominated by the expansion of $R_2 (
x )$ (see Appendix \ref{contas}). The contribution of the exchange interaction to the
term proportional $s^2$ is positive at low doping, and a continuous ferromagnetic 
transition is not possible. This contribution becomes negative only 
for $n=\delta A_0 \ge 0.059$.
As in the previous case, we can also analyze the system energy for large
values of the magnetization. 
We obtain an instability to a ferromagnetic state with full polarization
($m=2-2n$),
which for $n\ne 0$ leads to a state with both electron and hole carriers with
different Fermi surface areas. 
The dependence of the coupling constant $g_c$ on $n$ is given in
Fig.\ref{phase_diagram}
by the dashed line. 
(See more on the conclusions about a speculative
scenario for the origin of electrons and hole pockets in graphite.)

\subsubsection{Doped case, $n \neq 0$, two types of carriers
in the ferromagnetic phase} 

\begin{figure}[htf]
\begin{center}
\includegraphics*[width=8cm]{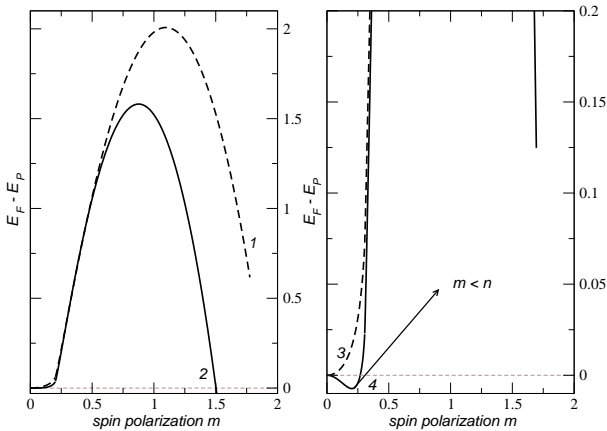}
\end{center}
\caption{
\label{phase_diag} Behavior of the energy curves as function
of the magnetization for the points marked in the phase diagram
of Fig.\ref{phase_diagram}.}
\end{figure}

In this case the calculation is analogous to the previous one.
The change in energy in this case is given by:
\begin{eqnarray}
\Delta E &= &\Delta K + \Delta E_{\rm ex} = \frac {A_0}{6\pi}v_F\hbar 
(\ku^3-\kd^3-2 \kf^3) \nonumber \\
&- &\frac{A_0}{(2 \pi )^2} \frac{e^2}{\eo}
\left[ \ku^3 R_1(1)+\kd^3 R_1(1)-2 \kf^3 R_1(1) 
 \right.
 \nonumber \\
&-&2 \kc \kd^2 R_1 \left( \frac{\kd}{\kc} \right) + 2 \kc \ku^2
 R_2 \left( \frac{\ku}{\kc} \right) \nonumber\\
&-&\left.4 \kc^2 R_2 \left( \frac{\kf}{\kc} \right) 
\right]\,.
\label{case_c}
\end{eqnarray}
As in the two previous cases, the leading term when $\kf , \ku , \kd \ll \kc$
is due to the expansion of the function $R_2 ( x )$, which leads to an
increase in the exchange energy, which is detrimental for ferromagnetism.
The energy change as a function of $m$ is shown in the right hand panel
of Fig.\ref{phase_diag}. We show the energy at points $3$ (paramagnetic) 
and $4$ (ferromagnetic) of Fig.~\ref{phase_diagram}. 
The transition in this case is second order with only partial polarization,
$m<n$. As a consequence only one type of carries exist.  The dependence of the
coupling  constant $g_c$ on $n$ is given in
Fig.\ref{phase_diagram} by the solid line.

\subsection{Gapful system}

A gap can open in the Dirac spectrum when the two sites in the unit cell of the
honeycomb lattice model become inequivalent equivalent. In this case,  the kinetic energy
Hamiltonian, Eq.(\ref{hkin}) changes to:
\begin{equation}
\Hkin ( \bm k ) \equiv \left( 
\begin{array}{cc} \Delta &\vf \hbar(k_x + i k_y) \\ \vf\hbar (k_y - k_y) &- \Delta
  \end{array} \right) \,,
\label{hkin_gap}
\end{equation}
which leads to the modified dispersion relation,
\begin{equation}
\ek = \pm \sqrt{\Delta^2 + ( \hbar \vf \vert {\bm k} \vert )^2}\,.
\label{dispersion_gap}
\end{equation}
For wavevectors such that $\hbar \vf \vert {\bm k} \vert \gg \Delta$ 
the energies and wavefunctions are  essentially the ones found in 
the absence of the gap, as discussed previously. If the filling is such that the Fermi
wavevector satisfies this conditions, but $\kf \ll \kc$ the analysis
presented earlier remains valid. At sufficiently low fillings, 
$\hbar \vf \vert \kf \vert \ll \Delta$, the dispersion relation,
Eq.(\ref{hkin_gap})  
can be approximated by:
\begin{equation}
\ek \approx \pm \Delta \pm \frac{( \hbar \vf \vert \bm k \vert )^2}
{2 \Delta}\,,
\end{equation}
and the bands depend quadratically on the wave vector and we 
can define an effective mass $m^* = \Delta/\vf^2$. Hence, the contribution
of the kinetic energy to the polarization energy is formally similar to that
obtained for an 2D electron gas with parabolic dispersion discussed extensively
in the literature. In this case, the spinor wave function becomes:
\begin{equation}
\Psi_{\bm k,\sigma} ( \bm r ) \simeq 
 \left( \begin{array}{c}  e^{i \bm k \bm r}
    \\ 0 
\end{array} \right) \chi_\sigma\,,
    \end{equation} 
for the upper sub-band, while the weight of the spinor is concentrated on
$\pb$, Eq.(\ref{wavefunction}), for the lower sub-band. This change modifies
significantly the spinor overlap factor in the calculation of the exchange
integral, Eq.(\ref{exchange_integral}). The overlap between Bloch states
  in different bands for momenta near the Fermi points vanishes (see the
  discussion at the end of Section III.A.1). These states do not give rise to
  inter-band contributions.  The only inter-band contributions which need to
be included are due to interactions between states far from the chemical
potential among themselves, and between these states at the bottom of the
lower band and those at the Fermi level. These terms are not modified when
the system is polarized, and they do not contribute to the exchange
instability.
The remaining intraband term is
equivalent to that derived for the electron gas with parabolic dispersion relation.  The change in
energy when the polarized state is formed can be written 
as
\begin{eqnarray}
\Delta E &=& \Delta K + \Delta E_{\rm ex} = \frac{A_0}{8 \pi} \frac{\vf^2}{2
  \Delta} ( \ku^4 + \kd^4 - 2 \kf^4 ) 
\nonumber\\
&-& \frac{A_0}{(2 \pi)^2} \frac{e^2}{\eo} 
\frac{4}{3} ( \ku^3 + \kd^3 - 2 \kf^3 )\,,
\end{eqnarray}
As in the usual case of the 2D electron gas, the system shows an
instability toward a ferromagnetic state when $\kf \le ( 16 \Delta e^2 ) / (
\pi \vf^2 \eo )$. In agreement with the previous discussion, this instability
vanishes when $\Delta \rightarrow 0$.

\subsection{The effect of disorder}

We approximate the effects of disorder on the average electronic structure by
means of the CPA \cite{S67}. This approximation describes
the effects of disorder on the electronic structure by means of a local self
energy, $\Sigma ( \omega )$ which is calculated self consistently.
While CPA cannot describe localization effects, it still gives very good
results for the physical properties of graphene \cite{nuno}. 

The total energy, including the exchange contribution, can be expressed in
terms of single particle Green's functions, which are calculated within the
CPA. The main steps of the calculation are sketched in Appendix B.
We assume that the disorder is induced by vacancies, as likely to occur in
samples treated by proton bombardment. The amount of disorder is parametrized
by the concentration o vacancies, $n_{\rm vac}$. The CPA leads to a density of
states which is finite at $\omega = 0$, and decays for $\omega \gg v_{\rm F}
n_{\rm vac}^{1/2}$\cite{nuno}.

Assuming that $\lim_{\omega   \rightarrow 0} {\rm Im} \Se = 
\So \sim ( \hbar \vf ) / l$, where $l$ is the
average distance between vacancies \cite{nuno} the calculations in Appendix
B admit some simplifications.  If the concentration of vacancies is
small, $\So \ll \hbar \vf | \kc |$. At large energies the CPA result vanishes quite fast
as a function of energy, $\lim_{\omega
  \rightarrow \pm \hbar \vf | \kc |} \Se = 0$. Disorder only changes
significantly the results obtained for a clean plane if $\ef \ll \So$. This
regime corresponds to electronic densities such that $| n | \ll n_0 = 
( \So / \hbar \vf )^2 / 2 \pi$.

 In this limit, we can approximately write
\begin{equation}
n_{\bf \vec{k}}^{\pm}  \approx \left\{ \begin{array}{lr} 0 &\hbar \vf
    | \bm k | \gg \So\,, \\ 1/2 + \frac{\ef}{\pi \So} & \hbar \vf
    | \bm k | \ll \So\,, \end{array} \right. \label{nk_approx} \, .
\end{equation}  
where the $\pm$ index refers to the two subbands of the noninteracting system
(see Appendix B).
   
The total density of carriers is obtained by integrating this expression over ${\bf
  \vec{k}}$ (see Appendix B). Finally, we can also calculate the density of
  states per unit area and unit energy, which, for $| \omega | \leq \So$,
  becomes a constant:
\begin{equation}
D(\omega) = D_0 \approx \frac{1}{2 \pi} \frac{\So}{\vf^2} \log \left(
  \frac{\hbar \vf \kc}{\So} \right) \,,
\hspace{0.5cm}
| \omega | , | \ef | \ll \So\,, 
\label{dos_disorder_2}
\end{equation}
A constant density of states implies that the total number of carriers scales
as $n \approx D_0 \ef$, instead of the relation $n \propto  \ef^2$ obtained for the
clean system.

From equations~(\ref{nk_approx}) and (\ref{dos_disorder_2}) we can infer that
both the kinetic energy and the exchange energy depend quadratically on the
density of carriers, since $K ( n ) - K ( 0 )$ and $E_{\rm exch} ( n ) -
E_{\rm exch} ( 0 )$ scale as  $ \ef^2 ( n ) \sim n^2$.
In addition, we know that for $n \approx n_0$
the values of $K ( n )$ and $E_{\rm exch} ( n )$ should be comparable to
those obtained in the absence of disorder. Then, we can write:
\begin{eqnarray}
K ( n ) & \approx &c_{\rm kin} \frac{2 A_0 \So^3}{3 \pi \hbar^2 \vf^2} \left( \frac{n}{n_0}
\right)^2, \nonumber \\
E_{\rm exch} ( n ) & \approx &- c_{\rm exch} \frac{A_0 e^2 \So^3}{3 \pi^2 \eo \hbar^3
  \vf^3} \left( \frac{n}{n_0} \right)^2,
\end{eqnarray} 
where $c_{\rm kin}$ and $c_{\rm exch}$ are numerical constants of order
unity. In a spin polarized system, we have:
\begin{widetext}
\begin{equation}
E_{\rm tot} ( n , m ) = \frac{1}{2} \left[ K ( n+m) + K (n-m) + E_{\rm exch}
  ( n+m ) + E_{\rm exch} (n-m) \right],
\end{equation}
so that:
\begin{equation}
\Delta E = \Delta K + \Delta E_{\rm exch} = 
c_{\rm kin} \frac{2 A_0 \So^3}{3 \pi \hbar^2 \vf^2} \left( \frac{m}{n_0}
\right)^2 - c_{\rm exch} \frac{A_0 e^2 \So^3}{3 \pi^2 \eo \hbar^3
  \vf^3} \left( \frac{m}{n_0} \right)^2 \, .
\end{equation}
\end{widetext}
The ferromagnetic phase is stable provided that:
\begin{equation}
g_{c,{\rm disorder}}=\frac{e^2}{\eo \hbar \vf} >  \frac{ 2 \pi c_{\rm kin} } {c_{\rm exch}}, 
\end{equation}
This result implies that, if $n \ll n_0$ the critical coupling is independent
of the amount of disorder. 

We have estimated the ratio $c_{\rm exch} / c_{\rm
  kin}$ performing numerically the calculation described in Appendix B for
  suficiently low carrier concentration and density of vacancies. We find:
\begin{equation}
g_{c,{\rm disorder}} = \frac{e^2}{\eo \hbar \vf} \simeq 3.8\, ,
\end{equation}
indicating that in the case of disorder ferromagnetism
is stabilized at a smaller value of the Coulomb interaction. Thus, we can
conclude that, at least in CPA, ferromagnetism will be enhanced when disorder
is present, in agreement with the experimental data \cite{esquinazi,disorder}.

The enhancement of the tendency towards ferromagnetism in the presence of
disorder is due to the increase in the density of states at low energies. The
existence of these states implies that a finite polarization can be achieved
with a smaller cost in kinetic energy, in a qualitatively similar way to the
Stoner criterium which explains itinerant ferromagnetism in the presence of
short range interactions.

\section{Discussion and Conclusions}
\label{conclusions}

We have analyzed the ferromagnetic instabilities induced by the
exchange interaction in a system where the electronic structure can be
approximated by the 2D Dirac equation, as it is the case for
isolated graphene planes.

In pure graphene we have found that, as a function of doping, a ferromagnetic
transition is possible when the coupling constant is sufficiently large.
Our findings are summarized in the zero temperature phase diagram presented
in Fig. \ref{phase_diagram}. In this figure we represent
the critical coupling $g_c$ as function of the doping $n$.
There are two different regions in the phase diagram.
For small doping, $n<0.2$ the transition is first order, leading to a
ferromagnetic phase with spin polarization $m=2-2n$ and two
types of carriers (electrons and holes). For 
doping larger than $n>0.2$ the transition becomes of second order
with a magnetization smaller than the doping $n$ and one type
of carrier (electrons or holes). The connection between the
magnetization and the carrier type is unique to the Dirac fermion problem. 
We should emphasize that our calculation for the Dirac fermion problem
is at the same level of the one performed by Bloch, and therefore it is 
to be expected that an exact solution of this problem will modify quantitatively
the phase diagram analyzed here. It is also worth remarking that the
electronic structure shown in panel of (c) Fig. \ref{cones} shows that, in
the ferromagnetic phase, a nominally half filled system has electron and hole
pockets. The existence of these pockets does not depend on the presence of
intarlayer coherence, however

We have also analyzed the effect of the exchange interaction in disordered
systems using the CPA.
A continuous transition into a ferromagnetic phase is possible, and the
coupling required for its
existence is reduced with respect to the clean case. This tendency can be
qualitatively explained by noting that the disorder leads to an increase of
the density of states at low energies, making the system more polarizable.
This explanation is rather general, and it should not depend on the way the
effects of disorder are approximated.

Finally, one would ask how our results can be translated for the experiments
in disordered graphite \cite{esquinazi,disorder}. If we naively
think of graphite as a stacking of isolated graphene planes we can estimate
the value of the coupling constant for graphite to be $g \sim 2.8$ (for
$\epsilon_0 \approx 1$) \cite{BCP88}, and therefore far away from the
ferromagnetic region (corresponding to the dotted line in Fig.\ref{phase_diagram}).
The presence of disorder will definitely bring the value of the critical coupling
to lower values and according to our calculations $g_{c,{\rm disorder}} \approx 3.8$
would put dirty graphite at the borderline of a ferromagnetic
instability. 

Nevertheless, the picture of graphene as a non-interacting stacking of graphene
planes is certainly incorrect. Because of the absence of screening, long-range
forces will play a major role, and the graphene planes will interact via van der Waals
interactions. The problem of ferromagnetism in graphite still depends on the
better understanding of the coupling between graphene planes. More work has
to be developed in order to understand the problem of ferromagnetism in graphite.
In any case,
our results here are valid for single graphene planes and it would be very
interesting to investigate whether graphitic devices 
\cite{Netal04,Zetal04,Zetal05,Betal04} studied recently can sustain
any form of ferromagnetism.

\section{Acknowledgments}
N.M.R.P and  F. G. are thankful to the Quantum Condensed Matter
Visitor's Program at Boston University.
A.H.C.N. was partially supported through NSF grant DMR-0343790.
N. M. R. Peres would like to thank 
Funda\c{c}\~ao para a Ci\^encia e Tecnologia for a sabbatical grant
partially supporting his sabbatical leave.

\appendix

\section{Calculation of the exchange integral}
\label{contas}

The three dimensional integral in Eq.(\ref{exchange_integral}) can be written
as a combination of integrals of the form:
\begin{equation}
R_n(a)=\int_0^{2\pi}d\alpha
\int_0^1xdx\int_0^1ydy\frac {{\rm sign }(n)-(-1)^n
\cos\alpha}{\sqrt{x^2+y^2a^2-2xya\cos\alpha}}\,,
\label{exch_int}
\end{equation}
where: $n=0,1,2$, ${\rm sign }(n)$ gives the sign of $n$ and 
 ${\rm sign }(0)=0$. The values of the functions $R_n(a)$, for $a=0$, are
$R_0(0)=0$ and $R_1(0)=R_2(0)=\pi$. We also have:
\begin{eqnarray}
R_0(1)&=&\frac 2 3 \left(
-2 + \pi(\ln 2+1/2) + 4{\cal C} - \pi(1+\ln 4)/2
\right)
\nonumber
\\
&\simeq& 1.109\,, 
\nonumber
\\
R_1(1)&=& 8/3+R_0(1)\simeq 3.776\,,
\end{eqnarray} 
where ${\cal C}\simeq 0.915966$ 
is the Catalan constant.

Assuming that $0\le a\le 1$ we define: 
\begin{equation}
R_n(a)=\int_0^{2\pi}d\alpha [{\rm sign }(n)1-(-1)^n
\cos\alpha]K(\alpha,a)\,,
\end{equation}  
where $K(\alpha,a)$ is given by:
\begin{widetext}
\begin{eqnarray}
K(\alpha,a)&=&
\frac 1 {3a^2} 
\left[-(1+a^3)+(1+a^2)\sqrt{1+a^2-2a\cos\alpha}-
(1+a^3)\cos\alpha\ln(1-\cos\alpha)-a^3\cos\alpha\ln a
\right.\nonumber\\
&+&\left.
\cos\alpha\ln(a-\cos\alpha+\sqrt{1+a^2-2a\cos\alpha}
)-a^3\cos\alpha\ln (1-a\cos\alpha+\sqrt{1+a^2-2a\cos\alpha})
\right]\,
\end{eqnarray}
\end{widetext}
This expression allows us to obtain the expansions:
\begin{eqnarray}
R_0(a)&\simeq&\frac {\pi} 3[ - a\ln a + S_0(a)]\,,\\
R_n(a)&\simeq&\frac {\pi} 3[3+(-1)^n a\ln a + S_n(a)]\,, 
\label{expan}
\end{eqnarray}
for $n=1,2$ and
\begin{eqnarray}
S_0(a)&=&\left(2\ln2-\frac 1 6\right)a- \frac 9 {80}a^3-
\frac {45}{1792}a^5
\nonumber
\\
&-& \frac {175}{18432}a^7\,,
\\
S_n(a)&=&-(-1)^n\left(2\ln2-\frac 1 6\right)a-
\frac 3 8a^2+
(-1)^n\frac 9 {80}a^3
\nonumber
\\
&-&\frac 3{64}a^4+
(-1)^n\frac {45}{1792} a^5-
\frac {15}{1024}a^6
\nonumber
\\
&+& (-1)^n\frac {175}{18432} a^7 \,.
\end{eqnarray}
Note that $R_1(a)-R_2(a)=2R_0(a)$ is always satisfied.

\section{Calculation of the exchange energy in the presence of disorder.}

We write the one-electron energies in the absence of interactions and
disorder as:
\begin{equation}
\epsilon_{\bm  {k}}^{\pm} = \pm v_{\rm F} ( {\bm  {k}} ) \, ,
\end{equation}
up to some cutoff $k_c$, where the two signs correspond to the two bands in the
electronic spectrum. Using the CPA, the one electron Green's function can be written as:
\begin{equation}
G^{\pm} ( {\bm  {k}} , \omega ) = \frac{1}{\omega - \Se - \epsilon_{\bm  {k}}^{\pm}
} \, .
\end{equation}

The occupancy of a given state at fixed chemical potential, $\epsilon_{\rm
  F}$, is:
\begin{equation}
n_{\bm  {k}}^{\pm} = \int_{- \omega_c}^{\epsilon_{\rm F}} \frac{1}{\pi} {\rm Im}
  G^{\pm} ( {\bm  {k}} , \omega ) d \omega \, ,
\end{equation}
where a frequency cutoff, $\omega_c$ is also defined.

The total number o electrons, $n$, and the kinetic energy can be written as:
\begin{eqnarray}
n &= &\sum_{\alpha=\pm} \frac{2}{\pi} \int_0^{k_c}  n_{| {\bm
   {k}}|}^{\alpha}  k d k \, , 
  \nonumber 
  \\
K &= &\sum_{\alpha=\pm} \frac{2}{\pi} \int_0^{k_c} \alpha\epsilon_{|{\bm  {k}}|} n_{|{\bm
   {k}}|}^{\alpha} k dk \, .
\end{eqnarray}
These one-dimensional integrals are calculated numerically.
Finally, the exchange energy is:
\begin{widetext}
\begin{equation}
E_{\rm exch} = - \frac{e^2}{4 \pi^4} \int d^2 {\bm  {k}}_1 \int d^2 {\bm
   {k}}_2 \frac{[ n^+ ( {\bm  {k}}_1 ) + n^- ( {\bm  {k}}_2 ) ]^2
+ [  n^+ ( {\bm  {k}}_1 ) - n^- ( {\bm  {k}}_2 ) ]^2 
\cos [ \phi ( {\bm  {k}}_1 ) - \phi ( {\bm  {k}}_2 ) ]}{| {\bm  {k}}_1
  -  {\bm  {k}}_2 |  } \, ,
\end{equation}
\end{widetext}
and:
\begin{equation}
\phi ( {\bm  {k}} ) = \arctan \left( \frac{k_y}{k_x} \right) \, .
\end{equation}
This expression can be reduced to a three-dimensional integral, which is
calculated numerically.

The total energy, $E_{\rm tot} ( n ) = K ( n ) + E_{exch} ( n )$, 
can be written as:
\begin{equation}
E_{\rm tot} ( n ) = E_{\rm tot} ( n_\uparrow ) + E_{\rm tot} ( n_\downarrow ) \, .
\end{equation}
The exchange instability towards ferromagnetism implies that:
\begin{equation}
E_{\rm tot} ( n / 2  - \delta n ) + E_{\rm tot} ( n / 2 + \delta n ) < 2 E_{\rm tot} (
n / 2 ) \, ,
\end{equation}
so that:
\begin{equation}
\left. \frac{\partial^2 E_{\rm tot}}{\partial n^2} \right|_{n/2} < 0 \, .
\end{equation} 

\bibliography{graphite0}

\end{document}